\documentclass[journal]{IEEEtran}
\IEEEoverridecommandlockouts
\usepackage{cite}
\usepackage{amsmath,amssymb,amsfonts}
\usepackage{algorithmic}
\usepackage{graphicx}
\usepackage{tabularx}
\usepackage{textcomp}
\usepackage[hyphens]{url}
\usepackage[hidelinks]{hyperref}
\usepackage{orcidlink}
\usepackage{makecell}
\begin{document}

\title{Unpacking Generative AI in Education: Computational Modeling of Teacher and Student Perspectives in Social Media Discourse}

\IEEEaftertitletext{%
\vspace{-2em}
\begin{center}
\small
\begin{minipage}{\columnwidth}
\centering
This is the original preprint version posted on arXiv prior to peer review. It has not been updated to reflect changes made during review. The final version of the paper is now published in \textit{IEEE Transactions on Computational Social Systems}. \textbf{The final, published version is copyrighted by IEEE and is available at:} https://doi.org/10.1109/TCSS.2025.3630587. \textbf{The citation to the published article is as follows:} 

P. DeVito et al., ``Unpacking Generative AI in Education: Computational Modeling of Teacher and Student Perspectives in Social Media Discourse'' in \textit{IEEE Transactions on Computational Social Systems}, doi: 10.1109/TCSS.2025.3630587.

\end{minipage}
\end{center}
\vspace{1ex}
}

\author{%
Paulina~DeVito\orcidlink{0009-0004-8636-4637}, %
Akhil~Vallala\orcidlink{0009-0008-8758-4040}, %
Sean~Mcmahon, %
Yaroslav~Hinda, %
Benjamin~Thaw, %
Hanqi~Zhuang\orcidlink{0000-0002-6490-731X}, %
Hari~Kalva\orcidlink{0000-0002-7165-5499}%
\thanks{This work is a preprint submitted to arXiv.}
\thanks{All authors are with the Department of Electrical Engineering and Computer Science, Florida Atlantic University, Boca Raton, Florida 33431 USA (e\mbox{-}mails: pdevito2019@fau.edu, zhuang@fau.edu, hkalva@fau.edu).}
\thanks{This material is based upon work supported by the National Science Foundation under Grant No. 2332306. Any opinions, findings, and conclusions expressed in this material are those of the authors and do not necessarily reflect the views of the National Science Foundation.}%
}

\maketitle
\markboth{DeVito \MakeLowercase{\textit{et al.}}: Unpacking Generative AI in Education: Teacher and Student Perspectives}{}


\begin{abstract}
Generative AI (GAI) technologies are quickly reshaping the educational landscape. As adoption accelerates, understanding how students and educators perceive these tools is essential. This study presents one of the most comprehensive analyses to date of stakeholder discourse dynamics on GAI in education using social media data. Our dataset includes 1,199 Reddit posts and 13,959 corresponding top-level comments. We apply sentiment analysis, topic modeling, and author classification. To support this, we propose and validate a modular framework that leverages prompt-based large language models (LLMs) for analysis of online social discourse, and we evaluate this framework against classical natural language processing (NLP) models. Our GPT-4o pipeline consistently outperforms prior approaches across all tasks. For example, it achieved 90.6\% accuracy in sentiment analysis against gold-standard human annotations. Topic extraction uncovered 12 latent topics in the public discourse with varying sentiment and author distributions. Teachers and students convey optimism about GAI’s potential for personalized learning and productivity in higher education. However, key differences emerged: students often voice distress over false accusations of cheating by AI detectors, while teachers generally express concern about job security, academic integrity, and institutional pressures to adopt GAI tools. These contrasting perspectives highlight the tension between innovation and oversight in GAI-enabled learning environments. Our findings suggest a need for clearer institutional policies, more transparent GAI integration practices, and support mechanisms for both educators and students. More broadly, this study demonstrates the potential of LLM-based frameworks for modeling stakeholder discourse within online communities.
\end{abstract}


\section{Introduction}
\label{sec:introduction}


Due to its widespread popularity, Generative AI's (GAI) presence in the classroom is now a question of ``how'' rather than ``if.'' With GAI tools such as ChatGPT, Copilot, Gemini, and Claude integrated into teaching and learning, the emphasis has shifted to how to appropriately use this technology. GAI has been welcomed in some contexts. For example, Khan Academy’s AI tool, Khanmigo, is used by over 65,000 students and teachers through a partnership with Microsoft \cite{rosenbaum_microsoft_2024, beatty_khan_2024}. Arizona State University (ASU) also partnered with OpenAI to bring ChatGPT Edu \cite{noauthor_bring_nodate} to support teaching and research on campus \cite{noauthor_arizona_nodate}. However, GAI in education has faced controversies. In 2023, \textit{Associated Press} reported that NYC public schools restricted access to ChatGPT to prevent cheating \cite{obrien_explainer_2023} but later lifted the ban \cite{rosenblatt_new_2023}. Several institutions followed suit, and the media continues to report concerns about GAI’s impact on student outcomes \cite{liu_future_2023, sinha_how_2023}. 

To navigate the difficulties of incorporating GAI in education, it is essential to understand the distinct opinions of teachers and students. Recent survey-based studies reveal a mix of optimism and apprehension. Concerns raised by both teachers and students include plagiarism, misinformation, and AI bias \cite{chan_ai_2023}. Some educators suggest modifying class assessments to allow GAI use rather than fully banning it \cite{smolansky_educator_2023}. Students overall appear to show more tolerance toward AI tools \cite{burkhard_student_2022, chan_will_2024}.


While such studies offer valuable insights, they are often constrained by scope and sample size. In contrast, social media provides rich, unfiltered perspectives on how AI is perceived in education. To contribute new insights, this study analyzes 1,199 Reddit posts and 13,959 top-level comments discussing GAI in education. We model the online educational community as a social system composed of main stakeholder groups, primarily teachers and students, which reveals divergent perspectives regarding GAI. We apply three methods—sentiment analysis, topic modeling, and author classification—to extract key takeaways from this discourse. Given the context-reliant nature of the Reddit discussions in our dataset, we compare classical natural language processing (NLP) methods against a prompt-based LLM framework to determine which approach produces the most reliable results. Our findings aim to inform educators, students, researchers, and policymakers by offering data-driven insight into GAI attitudes, supporting future directions for both K–12 and higher education. Specifically, we address the following research questions:

\begin{itemize}
    \item \textbf{RQ1:} How do frontier GPT models compare to traditional NLP methods for analyzing online social discourse?
    \item \textbf{RQ2:} What common trends and perspectives do teachers and students share regarding GAI?
    \item \textbf{RQ3:} How do sentiments toward GAI vary across topic trends between teachers and students?
    \item \textbf{RQ4:} What overlooked aspects of teacher and student views can this analysis uncover?
\end{itemize}


This paper is structured as follows: Section II reviews relevant literature on GAI in education and NLP techniques for social media studies in this area. Section III outlines our methodology. Section IV presents results. Section V concludes with key takeaways and future directions.


\section{Related Work}
\label{sec:background}

\subsection{Opportunities and Challenges of Integrating GAI in Education}

The advancement of GAI technologies, driven by the transformer architecture \cite{vaswani_attention_2017} and LLMs such as GPT \cite{brown_language_2020} and BERT \cite{devlin_bert_2019}, bring powerful GAI tools like ChatGPT, Copilot, Gemini, and Claude into educational settings. These tools demonstrate impressive capabilities, including personalized tutoring, grading automation, and writing assistance \cite{baidoo-anu_education_2023, alasadi_generative_2023, han_teachers_2024, bull_generative_2023, rouzegar_generative_2024}, and are increasingly adopted by students, especially those under academic pressures \cite{abbas_is_2024}. Educator awareness also plays a role in GAI tool adoption; tool exposure and AI literacy are now recommended goals for both students and teachers \cite{nazaretsky_teachers_2022, chiu_future_2024}.

However, GAI technology also raises concerns in academic contexts. One major issue is the threat to academic integrity, as students may use GAI tools to plagiarize and circumvent genuine understanding \cite{farrokhnia_swot_2024, cotton_chatting_2024}. Advanced GPT models, such as GPT-4, possess advanced multimodal reasoning abilities through self-reflection and chain-of-thought (COT) reasoning \cite{wei_chain--thought_2022}, which poses a threat to the authenticity of assessments \cite{susnjak_chatgpt_2024}. Detection systems like Turnitin and GPTZero suffer from high false positive rates and an inability to detect text altered by a student \cite{fowler_analysis_2023, ardito_contra_2023}. Institutions are cautioning educators to understand these limitations \cite{moorhouse_generative_2023}. Beyond academic integrity concerns, educator job displacement from AI automation is also a worry for some \cite{gocen_artificial_2021}, and scholars have warned of the potential for GAI to spread bias and misinformation. Ethical frameworks focused on data transparency and user consent will be essential for responsible AI integration in education \cite{berendt_ai_2020}.

As GAI continues to advance, its impact on education is becoming increasingly significant. This requires a reconsideration of how the technology should be integrated into education. Perspectives from both teachers and students are a prerequisite for developing strategies that leverage its benefits while minimizing its drawbacks in the classroom.

\subsection{Analyzing Educational Discourse Using NLP}

Understanding how teachers and students perceive GAI requires analyzing their experiences and viewpoints. Social media platforms offer unfiltered public opinion that is often absent in controlled survey settings. However, the informal and noisy nature of such data—containing slang, sarcasm, and domain-specific contexts—necessitates advanced NLP techniques for interpretation, such as the foundational techniques of sentiment analysis and topic modeling.

Sentiment analysis is the process of interpreting opinions in textual data. Classical lexicon-based models like VADER \cite{hutto_vader_2014} and TextBlob \cite{noauthor_textblob_nodate} are frequently applied in social media studies due to their simplicity and domain-agnostic design but often struggle to analyze complex language and mixed sentiments. Machine learning (ML) model approaches address some of these limitations but typically require substantial labeled data and feature engineering. Recent advancements in LLMs have catalyzed a new direction in this field. GPT and other LLMs have shown strong performance—even with just prompt-based zero-shot classification—in sentiment tasks across domains \cite{wang_is_2024, zhang_sentiment_2023, kheiri_sentimentgpt_2023},  outperforming traditional ML models on benchmarks like SemEval 2017 Task 4 \cite{rosenthal_semeval-2017_2019}. 

Topic modeling is a form of content analysis that identifies prevalent themes in a corpus. One of the most widely used topic modeling methods is Latent Dirichlet Allocation (LDA) \cite{blei_latent_2003}, which is a generative probabilistic model that treats each document as a mixture of topics \cite{tong_text_2016}. There are also other newer topic modeling methods, such as BERTopic \cite{grootendorst_bertopic_2022} and Top2Vec \cite{angelov_top2vec_2020}. Despite being popular, classical topic models like LDA face limitations such as a limited contextual understanding and requiring ``reading the tea leaves'' to interpret topics. To overcome these limitations, prompt-based LLM approaches are currently being explored for topic modeling as well \cite{phamTopicGPTPromptbasedTopic2024, wang_prompting_2023, mu_large_2024}. Collectively, these studies suggest that LLMs can complement or even outperform traditional methods in both sentiment and topic modeling tasks.

\begin{figure*}[htbp]
\centerline{\includegraphics[width=\textwidth]{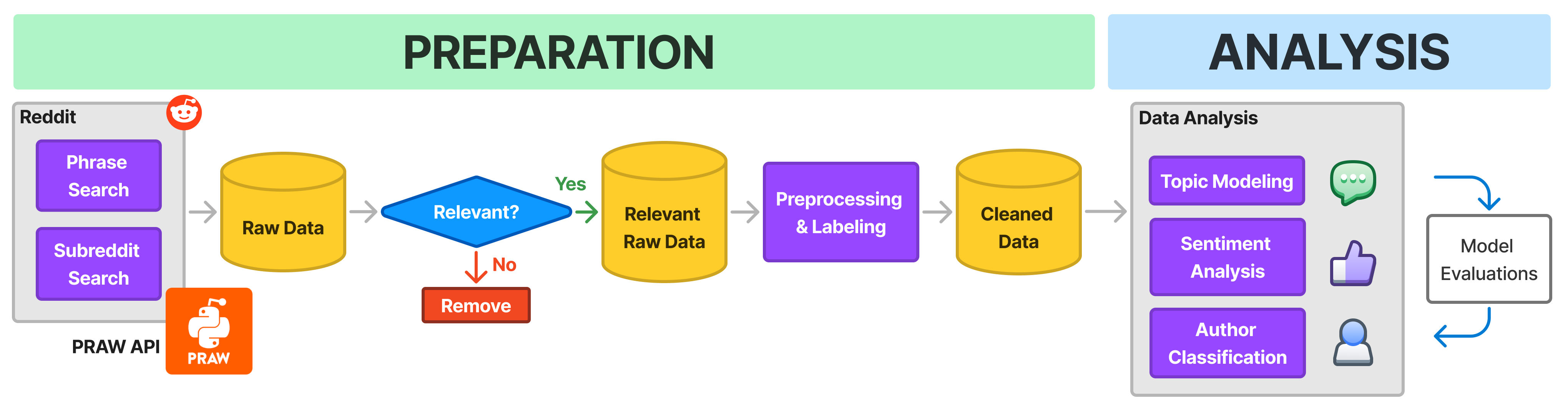}}
\caption{Overview of the study's data preparation and analysis pipeline.}
\label{fig:overview-flowchart}
\end{figure*}

\subsection{Positioning this Study in Relation to Prior Work}

Most social media studies continue to rely on established lexicon-based and ML methods due to their demonstrated effectiveness. For instance, some studies have implemented these classical sentiment analysis and topic modeling models to determine the public’s thoughts on online learning during the COVID-19 pandemic \cite{li_sentiment_2023, mujahid_sentiment_2021}. Since ChatGPT’s release in 2021, analyzing social media discourse about the tool, both generally and in educational contexts, has become a growing area of study. Prior work has revealed a mix of concerns and optimism toward ChatGPT, with most users engaging with the tool as a writing, coding, and research assistant \cite{choi_exploring_2023}. Another study found that negative perceptions of ChatGPT slightly declined after a year since its launch, and discussions covered a wide range of topics such as implications of ChatGPT in the job market, education, and software development \cite{naing_public_2024}. Though not their focus, these studies highlighted education as a key concern in ChatGPT discourse. 

Some studies using Twitter data have narrowed the focus to GAI in education. Fütterer et al. \cite{futterer_chatgpt_2023} analyzed 16.8 million tweets to examine early public opinion of ChatGPT just after release. Although the analysis initially covered broad topics with BERTopic, it later focused on education as a primary domain with generally positive sentiment toward ChatGPT in educational contexts. However, this study did not encompass other GAI tools and did not isolate teacher- or student-specific perspectives. The study of Kuhaneswaran et al. \cite{kuhaneswaran_exploring_2024} faced similar limitations. The researchers collected tweets related to the use of ChatGPT in education and applied LDA topic modeling, which showed a mix of wariness and positivity among post authors; however, they did not distinguish between teacher and student perspectives. Smith-Mutegi et al. \cite{smith-mutegi_perceptions_2025} conducted a sentiment analysis of tweets to explore public perceptions of AI in STEM education specifically, but their study did not extend to broader educational domains nor incorporate stakeholder role differentiation. Like many of these Twitter-based studies, the analysis relied on Twitter's short posts and traditional sentiment tools without benchmarking.

A few social media studies have narrowed their focus to educators only. Mamo et al. \cite{mamo_higher_2024} analyzed faculty tweets, reporting varying emotional distributions using lexicon tools. While useful, this study was limited to higher education faculty specifically. Closer in scope to our study, Lee \cite{lee_community_2024} focused on educators in Reddit’s r/Teachers subreddit. Lee used RoBERTa sentiment analysis and BERTopic topic modeling to analyze teacher reactions to ChatGPT. While notable for the use of Reddit and teacher-specific framing, this study was constrained to a single subreddit and lacked both an analysis of student perspective and benchmarking to assess the performance of the transformer-based models.

Oh et al. \cite{oh_explore_2024} provides the closest comparator to our study. Their study analyzed 3,400 Reddit comments relating to GAI in computer science (CS) education, applying both BERT-based and prompt-based sentiment analysis. However, their focus was restricted to only CS education. Their dataset included only comments, and their relevance filtering was based on keyword selection and LLM classification without manual validation. While they applied two distinct LLM-based sentiment models for comparison, they did not include benchmarking against human-labeled sentiment, topic modeling, and differentiation of author roles.

Building on prior research, our study offers several methodological and conceptual advancements:

\begin{itemize}
    \item We analyze a broad Reddit dataset including both posts and comments with manually validated relevance filtering.
    \item We apply prompt-based topic modeling using GPT-4o, requiring no fine-tuning, to derive more interpretable themes from the data. This addresses the limitations of traditional models like LDA.
    \item We evaluate classical sentiment tools (VADER, TextBlob) and multiple GPT model variants against gold-standard, human-labeled annotations.
    \item We classify authors by stakeholder group (e.g., student, teacher), enabling comparisons across roles.
\end{itemize}

To the best of our knowledge, such an extensive approach has not been fully explored in prior research within this domain. Our methodology is described in detail in the next section.


\section{Methodology}
\label{sec:methodology}

\subsection{Data Collection and Preprocessing}

Figure~\ref{fig:overview-flowchart} provides a high-level overview of the data preparation and analysis steps involved in the study. Data for this study was collected from Reddit, a forum-style social media platform where users create posts within communities called subreddits. Each post can receive comments, which form threaded discussions. Initial comment responses to the post are referred to as top-level or depth-0 comments. Top-level comments were the focus of this study, as they are typically more relevant to the original post and less likely to be off-topic conversational tangents. Users can upvote posts and comments to signal approval, which can influence visibility.  

\begin{table}[htbp]
    \caption{Keyword Phrases for Reddit Data Collection}
    \label{tab:keyword-phrases-reddit-data}
    \centering
    \begin{tabular}{ll}
        \hline
        \multicolumn{2}{c}{\textbf{Keyword Phrase}}  \\ \hline
        1  &  ``Generative AI in education''  \\ 
        2  &  ``AI teaching tools''  \\ 
        3  &  ``ChatGPT educational applications''  \\ 
        4  &  ``AI tutors in schools''  \\ 
        5  &  ``Generative AI impact on learning''  \\ 
        6  &  ``AI writing tools in education''  \\ 
        7  &  ``Educational AI technology trends''  \\ 
        8  &  ``AI and personalized learning''  \\ 
        9  &  ``Challenges of AI in education''  \\ 
        10 &  ``Future of education with AI''  \\ 
        11 &  ``Generative AI and academic integrity''  \\ 
        12 &  ``Using AI for educational content creation''  \\ 
        13 &  ``AI in classroom experiences''  \\ 
        14 &  ``Student engagement with Generative AI''  \\ 
        15 &  ``AI and educational equity''  \\ \hline
    \end{tabular}
\end{table}

Reddit data was collected over a six-month period in 2024 leveraging the PRAW (Python Reddit API Wrapper) library \cite{noauthor_praw_nodate}. The workflow was structured to collect Reddit data based on keyword phrases described in Table~\ref{tab:keyword-phrases-reddit-data} and across Reddit's post ranking criteria: “new,” “hot,” “top,” “rising,” “random\_rising,” “controversial,” and “gilded.” The script initially retrieved historical data and then continued to collect new data daily (“new” category) and weekly (all other categories). To supplement data potentially excluded by API rate limitations, we independently scraped the subreddits of origin from the collected Reddit posts, using the Table~\ref{tab:keyword-phrases-reddit-data} phrases once again to find more posts related to GAI in education. Duplicate posts and comments that emerged from these two simultaneous processes were removed.

Some posts containing one or more of the target keywords were irrelevant to the intended focus of this study. This stemmed from the nature of the keyword phrase searching with PRAW, which relies on keyword matching rather than contextual understanding to query Reddit posts. After carefully reading each post, the authors manually labeled them as either 0 (``irrelevant'') or 1 (``relevant''). Irrelevant posts were removed to ensure quality and validity of the final Reddit dataset.

Standard preprocessing was applied to the Reddit data. For posts, the title and body were combined, while only the body was used for comments. Text was cleaned by removing links, special characters, and entries from ``AutoModerator'' or with ``null'' authors. All text was lowercased, tokenized, stripped of stop words, and lemmatized for non-GPT tasks. Timestamps were also converted into a standardized format. 

After preprocessing, 1,199 Reddit posts and 13,959 Reddit comments remained for analysis. The top five subreddits where data was primarily sourced include r/ChatGPT, r/Professors, r/Teachers, r/ArtificialIntelligence, and r/ChatGPTautomation. There were 781 unique users among the 1,199 Reddit posts and 12,598 unique users among the 13,959 Reddit comments, indicating a diverse range of voices participating in the discussions. Moreover, there was strong engagement, with an average of 11.6 comments on each post.

\subsection{Topic Modeling}

To identify themes in the Reddit discussions, this study's topic modeling component leverages GPT-4o's advanced language understanding to extract latent topics from the online social discourse. We benchmark this against a classical NLP method, LDA \cite{blei_latent_2003}, which treats each document as a mixture of $k$ topics. To determine the optimal number of $k$ topics for LDA, coherence and perplexity scores were calculated using the Gensim library \cite{rehurek_software_2010}. 

GPT, a generative language model, fundamentally differs from LDA, a probabilistic model that infers topic distributions. However, GPT can be adapted for topic extraction via prompting. The rationale for exploring a GPT-based method lies in the characteristics of the Reddit dataset. Social media posts tend to contain noise (e.g., misspellings, slang), which can make LDA potentially less effective. Additionally, frequent but meaningful tokens complicate stop word removal. Excluding these tokens risks losing important word co-occurrences, but retaining them can blur topic boundaries with LDA. We expected GPT to mitigate these issues by utilizing contextual word embeddings and self-attention mechanisms to enhance topic differentiation. Specifically, GPT-4o was selected as the model of choice due to its strong language understanding capabilities. In fact, prior work has shown newer frontier GPT models performing better than older versions in a similar LLM topic modeling framework \cite{phamTopicGPTPromptbasedTopic2024}. Since this study emphasizes topic interpretability over predictive performance, direct comparisons between GPT models were less relevant than benchmarking GPT-4o against LDA.

\begin{table}
    \caption{Topic Modeling Prompting for GPT Models}
    \label{tab:topic-modeling-prompts}
    \centering
    \begin{tabularx}{\columnwidth}{>{\raggedright\arraybackslash}X}
        \hline
        \textbf{Reddit Post Prompts}  \\ \hline
        \texttt{\textbf{Prompt 1:} "\textit{Find the top words of the given post.}"}  \\ 
        \texttt{post = dspy.InputField(desc="\textit{A Reddit post.}")}  \\
        \texttt{top\_words = dspy.OutputField(desc="\textit{From the post, extract the top 10 words that best represent the main concepts and themes discussed. Focus on keywords that are central to the topic of the post. Don't format the words in a numbered list.}")}  \\ \hline
        \texttt{\textbf{Prompt 2:} "\textit{Cluster the list of top words into distinct topics.}"}  \\ 
        \texttt{top\_words\_list = dspy.InputField(desc="\textit{A list of top words taken from individual posts.}")}  \\ 
        \texttt{clustered\_topics = dspy.OutputField(desc="\textit{Group words that are semantically similar and relevant to the same theme together. Ensure that each cluster represents a unique topic. Don't list the top words in the output. Output a detailed title to each cluster and list the clusters numerically.}")}  \\ \hline
        \texttt{\textbf{Prompt 3:} "\textit{From the given list of topic clusters, label the the post topic.}"}  \\ 
        \texttt{top\_words\_list = dspy.InputField(desc="\textit{A list of top words from a post.}")}  \\
        \texttt{clustered\_topics = dspy.OutputField(desc="\textit{Assign the post to the most relevant cluster based on its main themes and concepts. None is not an option. Include the number with the associated topic, for example 'number here'. **test topic label**}")}  \\ \hline
    \end{tabularx}
\end{table}

The chained prompts used for the GPT approach are shown in Table~\ref{tab:topic-modeling-prompts}\footnote{DSPy, a recently released framework for LLM prompt engineering \cite{khattab_dspy_2023}, was utilized solely for code modularity. It was utilized for the sentiment analysis and author classification GPT tasks as well.}. Posts were first distilled into representative top words. This step reduced the length of the data within the context window while preserving necessary information. It also drastically lowered token usage, thereby reducing both computational cost and runtime complexity. A large list of top words from all the Reddit posts were then clustered into distinct topics. Finally, based on its representative top words, each post is associated with the topic cluster that best represents its textual content. Top-level comments inherited the same topic as their parent post.

\subsection{Sentiment Analysis}

The sentiment analysis component of the framework applies prompt-based GPT classification benchmarked against two traditional NLP models: VADER and TextBlob. Several GPT models were tested, namely GPT-3.5-turbo, GPT-4o-mini, and GPT-4o. The GPT models were instructed with the prompts shown in Table~\ref{tab:sentiment-analysis-prompts}.\footnote{The base prompts were augmented with examples for few-shot prompting tests.} The aim was to find a model that best represented the opinions of students, teachers, and other stakeholders. 

\begin{table}[htbp]
    \caption{Sentiment Analysis Prompting for GPT Models}
    \label{tab:sentiment-analysis-prompts}
    \centering
    \begin{tabularx}{\columnwidth}{>{\raggedright\arraybackslash}X}
        \hline
        \textbf{Reddit Post Prompt}  \\ \hline
        \texttt{"\textit{Classify the sentiment of the post. Assign a sentiment score from 0 to 2. 0 signifies the negative sentiment, 1 indicates neutral sentiment, and 2 positive sentiment.}"}  \\ 
        \texttt{post = dspy.InputField(desc="\textit{A Reddit post.}")}  \\
        \texttt{sentiment = dspy.OutputField(desc="\textit{Only output the sentiment score.}")} \\ \hline 
        \textbf{Reddit Comment Prompt}  \\ \hline
        \texttt{"\textit{Classify the sentiment of the comment based on the given post. Assign a sentiment score from 0 to 2. 0 signifies the negative sentiment, 1 indicates neutral sentiment, and 2 positive sentiment.}"}  \\ 
        \texttt{comment = dspy.InputField(desc="\textit{A Reddit comment discussing the post.}")} \\
        \texttt{post = dspy.InputField(desc="\textit{The post should act as the topic the comments are discussing.}")} \\
        \texttt{sentiment = dspy.OutputField(desc="\textit{Only output the sentiment score.}")}  \\ \hline
    \end{tabularx}
\end{table}

All models tested for sentiment analysis used a three-class classification: positive, negative, and neutral. In the hopes of obtaining optimal sentiment results and due to the brevity of the Reddit comments, posts were provided as context in determining comment sentiment with the GPT models. It should be noted that providing context this way is not possible with VADER and TextBlob unless the parent post is provided in the same document string as the comment, which could potentially skew results. For this reason, VADER and TextBlob were only provided the comment tokens in this study. 

Variations of each GPT model with zero-shot prompting and with few-shot prompting were tested. We expected that few-shot learning would allow the LLM to better generate the desired output on this study's Reddit dataset, especially since few-shot learning has been proven to yield better results than zero-shot in various contexts. Following a consensus-based manual review process, two authors of this study simultaneously assigned sentiment labels to Reddit posts and comments. Any disagreements in labels were resolved, and a final label was assigned after discussion by the labelers. Every Reddit post ($n_{posts}$=1,199) was manually labeled for sentiment. Given the large size of the Reddit comment dataset, a random set of comments was selected for labeling. This was done to strike a balance between feasibility of labeling and representativeness in the model comparisons. Therefore, 2,500 Reddit comments were chosen for manual sentiment labeling, and after data cleaning, 2,443 of 13,959 Reddit comments remained. Given the intensive nature of manual annotation, the labeling of posts and comments required substantial time and coordination. These ground truth labels enabled the calculation of accuracy and F1-score metrics for each model.

\subsection{Authorship Classification}

Author classification was the last task performed on the Reddit dataset. The GPT models were given the prompt shown in Table~\ref{tab:author-classification-prompts}.\footnote{The base prompts were augmented with examples for few-shot prompting tests.} Similar to sentiment analysis, we tested GPT-3.5-turbo, GPT-4o-mini, and GPT-4o, all with and without few-shot prompting. In addition to categorizing authors as ``students,'' ``teachers,'' or ``others,'' we also introduced a distinct ``parent'' category. This decision was made after observing a small but notable number of posts authored by individuals explicitly identifying as parents of students. Although parents represented a small portion of the dataset, we included them as a distinct author category to assess their presence in the discourse—an angle not typically examined in prior literature on generative AI in education.

\begin{table}
    \caption{Author Classification Prompting for GPT Models}
    \label{tab:author-classification-prompts}
    \centering
    \begin{tabularx}{\columnwidth}{>{\raggedright\arraybackslash}X}
        \hline
        \textbf{Reddit Post Prompt}  \\ \hline
        \texttt{"\textit{Classify the author of the post. Available labels are student, teacher, parent, or other.}"}  \\ 
        \texttt{post = dspy.InputField(desc="\textit{A Reddit post.}")}  \\ 
        \texttt{author = dspy.OutputField(desc="\textit{Please only output one author and only one word. If it is unclear, choose other.}")}  \\ \hline
        \textbf{Reddit Comment Prompt}  \\ \hline
        \texttt{"\textit{Classify the author of the comment. Available labels are student, teacher, parent, or other.}"}  \\ 
        \texttt{comment = dspy.InputField(desc="\textit{A Reddit comment.}")}  \\ 
        \texttt{author = dspy.OutputField(desc="\textit{Please only output one author and only one word. If it is unclear, choose other.}")}  \\ \hline
    \end{tabularx}
\end{table}

The goal of author classification with the GPT models was to identify students, teachers, parents, and others within the Reddit dataset. Once again, a consensus-based manual review process was conducted by two authors of this study, this time to assign author labels to every Reddit post ($n_{posts}$=1,199). It is important to clarify that authorship classification within the context of this study can be subjective, as both GPT and manual labeling could be incorrect due to Reddit’s anonymous nature. We adhered strictly to Reddit's privacy guidelines and made no attempts to de-anonymize or identify individual users. Human annotators followed a conservative labeling approach by labeling posts with ambiguous authorship as “other” rather than inferring roles without sufficient evidence. With that in mind, the performance of the GPT models should be interpreted as a reflection of their ability to infer author affiliations solely based on contextual cues within the text. 

\section{Results and Discussion}

This section presents the evaluation results of the proposed LLM-based computational framework across the three core tasks: topic modeling, sentiment analysis, and author classification. Model performance is evaluated against classical NLP methods, and patterns in the analyzed discourse are discussed.

\subsection{Model Comparisons}

For topic modeling, the results from LDA were compared with those from GPT-4o. For LDA, both $k$=4 and $k$=13 topics were evaluated because they represented local peaks in coherence score (around 0.39 for $k$=4 and 0.38 for $k$=13) while balancing relatively low perplexity. However, as can be seen by the top words for $k$=4 topics provided in Table~\ref{tab:lda-results}, LDA struggled to generate distinguishable, interpretable topics. 

\begin{table}[htbp]
    \caption{LDA $k$=4 Topics and Top 20 Words Per Topic}
    \label{tab:lda-results}
    \centering
    \begin{tabular}{m{1.25cm}m{6.75cm}}
        \hline
        \textbf{Topic 1}  &  ai student like use school chatgpt education would help using work teacher get think way could one learning people feel  \\ \hline
        \textbf{Topic 2}  &  ai student use using essay chatgpt would work like teacher class paper tool writing assignment one school time know used  \\ \hline
        \textbf{Topic 3}  &  ai learning student education chatgpt tool experience link technology educational language personalized future data also need use potential like human \\ \hline
        \textbf{Topic 4}  &  ai student writing tool academic knowledge use using like content human plagiarism technology text teacher generated essay work education detection  \\ \hline
    \end{tabular}
\end{table}

To numerically evaluate the difference between LDA and GPT-4o, we employed both intra-topic similarity (capturing semantic tightness within topics) and inter-topic distance (measuring separation between topics), calculated with two sentence embedding models (\texttt{all-MiniLM-L6-v2} \cite{noauthor_all-minilm-l6-v2_2024} and the much larger \texttt{all-mpnet-base-v2} \cite{noauthor_all-mpnet-base-v2_2024}) to mitigate embedding-specific biases. Given that LDA is a probabilistic model that infers latent topics from word distributions whereas GPT-4o is a generative language model that generates topics via prompt instructions, these metrics provide a model-agnostic framework for comparison between these fundamentally different methods. GPT-4o consistently outperformed LDA across both intra-topic similarity and inter-topic distance calculations for both embedding models. On the \texttt{all-mpnet-base-v2} embedding model, GPT-4o achieved an intra-topic similarity of 0.4431, which is a 10.1\% improvement over the best LDA configuration (4-topic, 0.4025). Additionally, GPT-4o demonstrated a higher inter-topic distance of 0.2134, which is a 70.2\% improvement over the best LDA configuration (13-topic, 0.1254). While the absolute inter-topic distance (0.2134) may appear modest, this reflects the lexical overlap in the Reddit discussions, where more frequently used terms often appeared across multiple distinct topics. 

These metrics show that GPT-generated topics were both internally more cohesive and significantly more distinct from one another. Furthermore, manual inspection confirmed that GPT-4o's topics were substantially more human-interpretable compared to those generated by LDA. This reinforces GPT-4o's superior capacity to capture nuanced structures in the text. Consequently, the topic clusters from GPT-4o were selected for analysis. 12 topics were identified, shown in Table~\ref{tab:topics}. 

For the sentiment analysis and author classification tasks, model comparisons using accuracy and F1 score metrics were performed. Sentiment analysis evaluation was conducted on both Reddit posts and comments; however, for space efficiency, we present detailed evaluation results only for comments, shown in Table~\ref{tab:model-comparison-sentiment-comments}. Performance patterns were consistent across both posts and comments: VADER and TextBlob performed rather poorly. The context-dependent nature of the Reddit discussions made it difficult for lexicon-based models to interpret sentiment accurately. These models also struggled with mixed opinions, leading to misclassifications. In stark contrast, GPT-4o with few-shot prompting achieved the highest accuracy and F1-scores across all sentiment classes for both posts and comments. Notably, this performance was obtained without any fine-tuning, suggesting that in-context learning may be sufficient for sentiment analysis of Reddit social media discourse.

\begin{table}[htbp]
    \caption{Model Comparison for Sentiment Analysis on Reddit Comments}
    \label{tab:model-comparison-sentiment-comments}
    \centering
    \begin{tabular}{lcccc}
        \hline
        \makecell{\textbf{Model}} & \makecell{\textbf{Accuracy}} & \makecell{\textbf{Positive} \\ \textbf{F1}} & \makecell{\textbf{Negative} \\ \textbf{F1}} & \makecell{\textbf{Neutral} \\ \textbf{F1}} \\ \hline
        4o FS & \textbf{90.63\%} & \textbf{91.91\%} & \textbf{94.06\%} & \textbf{82.55\%} \\ 
        4o ZS & 84.57\% & 86.70\% & 89.14\% & 73.23\% \\ 
        4o-mini FS & 84.20\% & 86.91\% & 89.41\% & 68.81\% \\ 
        4o-mini ZS & 84.49\% & 87.20\% & 90.11\% & 67.98\% \\ 
        3.5-turbo FS & 75.52\% & 80.35\% & 84.03\% & 40.55\% \\ 
        3.5-turbo ZS & 74.17\% & 79.10\% & 83.48\% & 35.64\% \\ 
        VADER & 60.17\% & 63.40\% & 66.95\% & 40.61\% \\ 
        TextBlob & 42.20\% & 46.18\% & 41.01\% & 40.73\% \\ \hline 
    \end{tabular}
    \begin{minipage}{7.25cm}
        \vspace{1ex}
        \raggedright
        \footnotesize
        Abbreviations: ``4o'' = GPT-4o; ``4o-mini'' = GPT-4o-mini; ``3.5-turbo'' = GPT-3.5-turbo. FS = Few-Shot; ZS = Zero-Shot.
    \end{minipage}
\end{table}

Identifying neutral sentiments in comments was most challenging for the GPT models tested. For instance, many posts in our dataset solicit guidance after an innocent student has been accused of using GAI. A typical post may read: \textit{“I’ve been accused of using ChatGPT on my paper. What should I do?”} Responses often include practical advice, such as: \textit{“Use Google Drive; it saves your revision history as you write. Thankfully, I wrote my essays like this.”} Across all GPT configurations, comments like these were frequently misclassified as positive. However, since they provide guidance rather than expressing a clear opinion about GAI, our annotators categorized them as neutral. While advice-giving may carry implicitly prosocial tones, it does not necessarily reflect a sentiment on GAI itself. Consequently, model performance was evaluated based on alignment with this criteria. Few-shot prompting substantially improved model performance by explicitly demonstrating how advice-oriented and emotionally inconclusive comments should be labeled as neutral.

Table~\ref{tab:model-comparison-author-posts} illustrates that GPT-4o with few-shot performed the best on the Reddit dataset for post author classification across all metrics. A common trend of error observed across every GPT variant was their more ``generous'' labeling than the human labels. For instance, if a post is structured around teaching, it might be classified under the ``teacher'' category, despite insignificant contextual proof of this. Consider this paraphrased example from our dataset: \textit{``Grading students on in-class work could make it harder for them to cheat.''} Although this example references student behavior, it does not clearly indicate the author's role. GPT might classify the author as a teacher based on the framing, whereas our more strict human annotators labeled it as ``other'' due to insufficient evidence. A teacher might have more explicitly contextualized the remark, e.g., by writing, \textit{“I feel I should start requiring in-class assignments to reduce cheating.”} It is important to reiterate that author classification for this study is inherently subjective, given Reddit’s anonymity. We once again note the following limitation: both model- and human-based labeling are susceptible to uncertainty when author role signaling is subtle or absent. 

\begin{table}[htbp]
    \caption{Model Comparison for Author Classification on Reddit Posts}
    \label{tab:model-comparison-author-posts}
    \centering
    \begin{tabular}{lccccc}
        \hline 
        \makecell{\textbf{Model}} & \makecell{\textbf{Accuracy}} & \makecell{\textbf{Teacher} \\ \textbf{F1}} & \makecell{\textbf{Student} \\ \textbf{F1}} & \makecell{\textbf{Parent} \\ \textbf{F1}} & \makecell{\textbf{Other} \\ \textbf{F1}} \\ \hline
        4o FS & \textbf{87.32\%} & \textbf{85.71\%} & \textbf{93.65\%} & \textbf{83.33\%} & \textbf{80.17\%} \\ 
        4o ZS & 80.07\% & 78.81\% & 90.84\% & 63.83\% & 65.21\%\\ 
        4o-mini FS & 77.48\% & 77.88\% & 89.55\% & 62.96\% & 57.04\% \\ 
        4o-mini ZS & 77.31\% & 77.15\% & 90.05\% & 59.65\% & 56.59\% \\
        3.5-turbo FS & 77.56\% & 77.35\% & 87.41\% & 60.87\% & 62.24\% \\ 
        3.5-turbo ZS & 75.40\% & 76.15\% & 87.43\% & 60.87\% & 58.43\% \\ \hline
    \end{tabular}
    \begin{minipage}{8.5cm}
        \vspace{1ex}
        \raggedright
        \footnotesize
        Abbreviations: ``4o'' = GPT-4o; ``4o-mini'' = GPT-4o-mini; ``3.5-turbo'' = GPT-3.5-turbo. FS = Few-Shot; ZS = Zero-Shot.
    \end{minipage}
\end{table}
\begin{table*}
    \caption{Topics from GPT-4o Topic Modeling Process}
    \label{tab:topics}
    \centering
    \begin{tabular}{>{\centering\arraybackslash}m{1cm} >{\centering\arraybackslash}m{3.25cm} m{12.25cm}}
        \hline
        \textbf{Topic \#} & \textbf{Topic Name} & \textbf{Counts and GPT-4o Topic Description}  \\ \hline
        1 & Personalized Learning & \textbf{($n_{posts}$=216, $n_{comments}$=1059)} This cluster includes words related to the integration of AI in educational settings, personalized learning experiences, and the use of AI tools to enhance teaching and learning. It covers aspects like intelligent tutoring systems, adaptive learning, and personalized feedback. \\ \hline
        2 & Ethics \& Integrity & \textbf{($n_{posts}$=367, $n_{comments}$=7119)} This cluster focuses on the ethical implications of using AI in education, issues of academic integrity, plagiarism detection, and the challenges of maintaining honesty in academic work. It includes discussions on ethical guidelines, policies, and the impact of AI on academic misconduct. \\ \hline
        3 & Tools \& Tech & \textbf{($n_{posts}$=173, $n_{comments}$=2077)} This cluster encompasses various AI tools and technologies used in educational contexts, such as ChatGPT, Grammarly, and other AI-driven applications. It includes the use of AI for writing assistance, content generation, and educational software. \\ \hline
        4 & Teacher Impact & \textbf{($n_{posts}$=37, $n_{comments}$=843)} This cluster addresses the implications of AI on the teaching profession, including the potential for AI to replace or assist teachers, the role of AI in classroom management, and the impact on teacher-student interactions. \\ \hline
        5 & Student Experience & \textbf{($n_{posts}$=58, $n_{comments}$=406)} This cluster includes words related to the student experience with AI, such as the use of AI for assignments, exams, and learning support. It covers the benefits and drawbacks of AI from the student perspective, including issues of stress, anxiety, and academic performance. \\ \hline
        6 & Policy & \textbf{($n_{posts}$=23, $n_{comments}$=196)} This cluster focuses on the policies and guidelines surrounding the use of AI in education. It includes discussions on institutional policies, government regulations, and the role of policymakers in shaping the future of AI in education. \\
        \hline
        7 & Higher Ed \& Research & \textbf{($n_{posts}$=149, $n_{comments}$=1014)} This cluster includes words related to the use of AI in higher education and academic research. It covers topics like AI-driven research tools, academic writing, and the impact of AI on university-level education and research practices. \\
        \hline
        8 & Language Learning & \textbf{($n_{posts}$=28, $n_{comments}$=112)} This cluster focuses on the use of AI in language learning, including tools for translation, language practice, and personalized language education. It includes discussions on the impact of AI on language acquisition and teaching methodologies. \\
        \hline
        9 & Student Assessment & \textbf{($n_{posts}$=48, $n_{comments}$=143)} This cluster includes words related to the use of AI for student assessment, grading, and feedback. It covers the use of AI to evaluate student work, provide personalized feedback, and the challenges of AI-driven assessment methods. \\
        \hline
        10 & Educational Content Creation & \textbf{($n_{posts}$=26, $n_{comments}$=113)} This cluster focuses on the use of AI for creating educational content, such as lesson plans, study materials, and interactive learning experiences. It includes discussions on the role of AI in content generation and the impact on educational resources. \\
        \hline
        11 & Educational Equity & \textbf{($n_{posts}$=3, $n_{comments}$=0)} This cluster addresses the role of AI in promoting educational equity, including access to personalized learning, bridging the digital divide, and ensuring that AI tools are accessible to all students regardless of socioeconomic background. \\
        \hline
        12 & Future of Education & \textbf{($n_{posts}$=71, $n_{comments}$=877)} This cluster includes words related to the future of education with AI, including predictions, trends, and the potential for AI to revolutionize educational practices. It covers discussions on the long-term impact of AI on the education system and the evolving future of education. \\
        \hline
    \end{tabular}
\end{table*}

\subsection{Topic Analysis}

\begin{figure}[htbp]
\centerline{\includegraphics[width=\columnwidth]{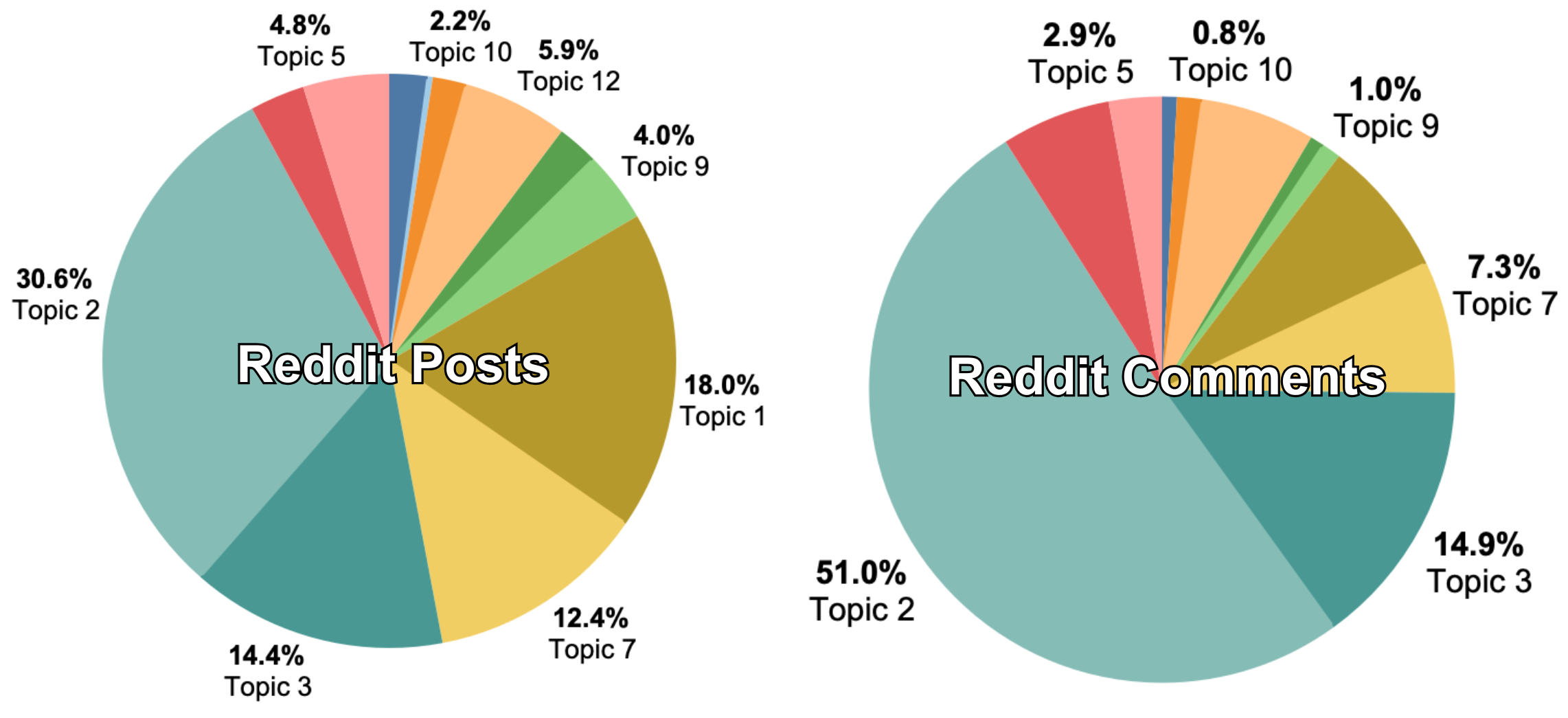}}
\caption{Distribution of the 12 topics among Reddit posts and comments.}
\label{fig:topic-distributions}
\end{figure}

\begin{figure*}[htbp]
\centerline{\includegraphics[width=\textwidth]{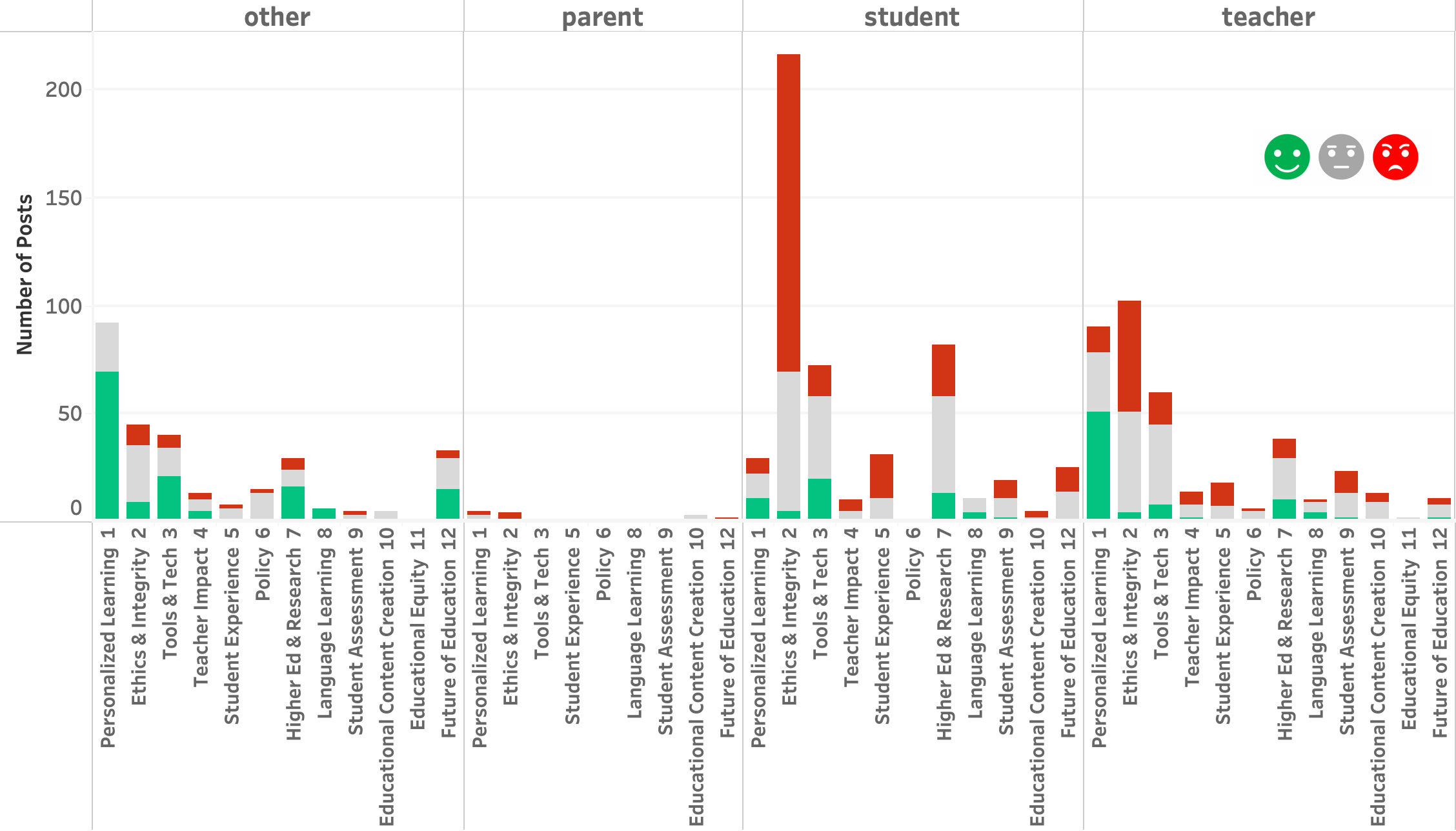}}
\caption{Combined analysis on the Reddit posts based on GPT-4o topic modeling, sentiment analysis, and author classification outputs. Green represents positive sentiment, gray represents neutral sentiment, and red represents negative sentiment.}
\label{fig:combined-analysis-posts}
\end{figure*}

\begin{figure*}[htbp]
\centerline{\includegraphics[width=\textwidth]{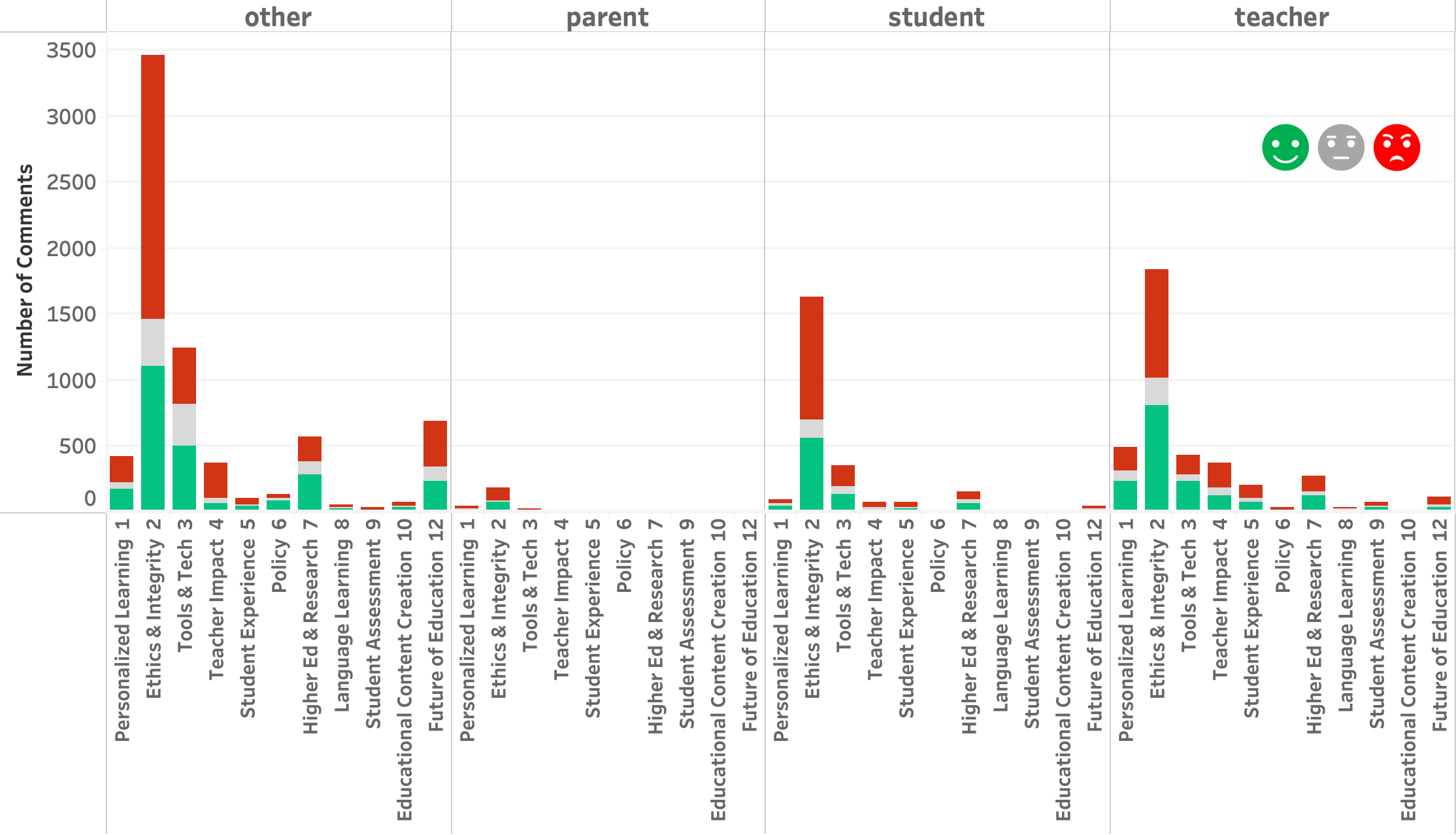}}
\caption{Combined analysis on the Reddit comments based on GPT-4o topic modeling, sentiment analysis, and author classification outputs. Green represents positive sentiment, gray represents neutral sentiment, and red represents negative sentiment.}
\label{fig:combined-analysis-comments}
\end{figure*}

Figure~\ref{fig:topic-distributions} displays the distribution of the 12 topics across Reddit posts and comments. Figures~\ref{fig:combined-analysis-posts} and~\ref{fig:combined-analysis-comments} detail the sentiments across each of the 12 topics and authors for Reddit posts and comments, respectively. These figures serve as a basis for the per-topic analysis and interpretation. To explore interaction dynamics, we observed the sentiment agreements between posts and their comments. Of the 13,959 comments, 6,308 (45.2\%) matched the sentiment of their parent post. This agreement rate suggests that Reddit discourse around GAI in education is rich in dialogue rather than echo-chambered because users more often reply with differing emotional perspectives. 

\textbf{Topic 1 (Personalized Learning)} focuses on personalized learning with GAI in education. It is the second-largest cluster, comprising 18\% of all posts. Overall, each author type leans positively toward this topic in both the posts and comments. Negative-to-positive sentiment ratios reflect this trend, with students (44:64) and teachers (187:291) expressing more positive than negative views. Many teachers, students, and others agree that GAI is valuable for tutoring and facilitating learning experiences from home. Teachers note that GAI helped them answer students' more complex questions and personalize their interactions. However, some authors with a negative sentiment argue that AI-driven personalized learning could undermine teachers' roles. Some teachers, in both posts and comments, express concerns that students are misusing AI instead of engaging in academically honest learning. Neutral sentiments state benefits of GAI but remain skeptical about its limitations.

\textbf{Topic 2 (Ethics \& Integrity)} accounts for the largest share of posts and comments in the dataset: 30.6\% and 51.0\%, respectively. An overwhelmingly large amount of posts and comments discuss cheating with GAI. A chi-square test reveals a significant association between sentiment and author role ($\chi^2(6)$ = 52.41, $p$ $<$ 0.0001), with students contributing disproportionately more negative posts on this topic compared to other groups. This pattern is also reflected in the students' negative-to-positive sentiment ratio (1073:569), which leans heavily negative for this topic. Because the majority of emotionally-driven discussion is concentrated in this topic, an additional detailed analysis was conducted and is presented in the \textit{Academic Integrity} section of this paper.

\textbf{Topic 3 (Tools \& Tech)} focuses on the use of task-specific tools (e.g., Midjourney) in educational contexts. Some teachers express that GAI tools are a great help, while others feel that these tools are not producing sophisticated outputs for their use cases. 

\textbf{Topic 4 (Teacher Impact)} has most posts express negative or neutral sentiments. Overall, both students (50:21) and teachers (192:129) show more negative than positive sentiment on this topic. Negative sentiments detail teachers' concerns about the shift in education or eventually being replaced by AI, while neutral posts consist of teachers wondering how they could boost their productivity with AI. One post in this topic gained significant attention, with over 8,000 upvotes and many commenters divided on the idea of justifying AI to automate grading while simultaneously not allowing students to use AI in the classroom.

\textbf{Topic 5 (Student Experience)} includes many mentions of teachers or students feeling that students are not using AI correctly and are simply copying from it. This is reflected in the high negative-to-positive ratios for students (56:39) and teachers (111:84).

\textbf{Topic 6 (Policy)} is smaller in scope. Most of the discussion centers around how educational institutions are attempting to integrate AI policies. However, these institutions are either implementing strict rules or facing hurdles due to the many considerations that need to be addressed.

\textbf{Topic 7 (Higher Ed \& Research)} is a larger topic and has very unique insights. Posts with negative sentiments typically detail undergraduate and graduate students misusing AI tools for research papers or other written work (e.g., discussion forums). However, many positive and neutral posts point out that GAI could enhance their study and research efficiency. Uniquely, some educators also express a willingness to educate students and fellow faculty about GAI technology within this topic. 

\textbf{Topic 8 (Language Learning)}, a smaller topic overall, includes teachers that express a generally positive view of GAI for language learning, with several authors claiming that GAI language learning tools are useful for teaching different languages, and others stating how they have used AI because they do not speak English as their first language. Comments on the latter posts tend to be uplifting and in support of using AI for this purpose. Students are either neutral or positive, with some mentioning they have used AI tools for learning other languages.

\textbf{Topic 9 (Student Assessment)} highlights neutral discussions among teachers considering the use of GAI for grading, often due to concerns that traditional grading demands detract from lesson creation—a pattern also mentioned in Topic 4 (Teacher Impact). Teachers also exhibit negative sentiments related to challenges with academic misconduct in assessments. Students in this cluster have generally neutral and negative sentiments, seeking practical guidance on using GAI tools for self-assessment while also reporting instances of false AI-related accusations in assessments. This is parallel to issues raised in Topic 2 (Ethics \& Integrity). 

\textbf{Topic 10 (Educational Content Creation)} discusses the use of GAI tools to create educational content, such as GAI-generated images for class assignments. Posts and comments reflect mixed and negative opinions on the effectiveness of GAI tools for these tasks. A small number of parent-authored posts appear in this topic, primarily focused on using GAI to assist with homeschooling. 

\textbf{Topic 11 (Educational Equity)} is the smallest topic, consisting of just three posts that focus on concerns about access and affordability of GAI tools in education. These posts address a digital divide and state that GAI technologies are not equally accessible to all students. While the discussion of equity is also embedded in other topics, such as Topic 6 (Policy), the posts in this topic appear appropriately clustered together due to their narrower rhetoric. The appearance of this small topic may reflect both a distinct semantic grouping and some degree of noise; however, this topic highlights that equity remains an important area of discussion within a subset of the data.

\textbf{Topic 12 (Future of Education)} includes posts that convey concerns about an AI revolution in education. Negative sentiments emphasize the potential erosion of critical thinking, the devaluation of certain majors (e.g., illustration), and job displacement for soon-to-graduate students. Teacher (58:45) and student (33:19) sentiment negative-to-positive ratios align with these worries. While some Reddit users share these concerns, others with neutral and positive sentiments argue that GAI’s current limitations prevent an immediate revolution. Positive posts and comments focus on the long-term benefits AI could offer to future learners, such as the ability to have a 24/7 tutor without a financial barrier. 

When aggregating sentiments across all posts and comments, we found that 46.9\% are negative, 37.7\% positive, and 15.4\% neutral. This skew toward negativity is closely linked to several high-volume topics in the dataset. In particular, Topic 2 (Ethics \& Integrity) dominated the negative discourse surrounding academic integrity and student experiences with AI misuse or false accusations. Topics like Topic 4 (Teacher Impact) and Topic 12 (Future of Education) also exhibited substantial negative sentiment, with many teachers expressing fears about job displacement and uncertainty regarding AI’s long-term role in education. In contrast, Topics such as Topic 1 (Personalized Learning) and Topic 7 (Higher Ed \& Research) contributed more positively to the discourse. However, these sentiments were insufficient in outweighing the emotionally-charged negative conversations.

\subsection{Academic Integrity}

Given the extensive discussion of academic integrity identified in Topic 2, we conducted another detailed manual review of each Reddit post to categorize the specific types of academic misconduct mentioned. This led to a significant discovery: approximately 25\% of all posts and 50\% of all comments\footnote{The number of associated comments ($n_{comments}$) is reported to illustrate engagement levels but was not used for this category classification, which was based on the content of original posts.} fall into one of four distinct categories, as listed below.

\begin{enumerate}
    \item \textbf{Student who was falsely accused of AI cheating} ($n_{posts}$=108 with $n_{comments}$=4893, approximately 90\% of posts have a negative sentiment)
    \item \textbf{Student who cheated with AI and was caught} ($n_{posts}$=33, $n_{comments}$=126, approximately 75\% of posts have a negative sentiment)
    \item \textbf{Student or fellow students who cheated with AI but were not caught} ($n_{posts}$=46, $n_{comments}$=285, a mix of negative and neutral/mixed sentiments for posts)
    \item \textbf{Teachers who believe they found cheating among a student or students} ($n_{posts}$=118, $n_{comments}$=1436, mostly negative sentiments for posts)
\end{enumerate}

In Category 1 posts, students share detailed experiences of being falsely accused of using AI to complete their assignments, which were typically writing assignments. Four out of the top five posts in terms of upvotes came from this category, indicating a high level of shared engagement in these conversations. Many comments on these posts offer advice to the student authors on how to handle these situations. Nearly all posts in this category convey a negative sentiment, which is a reflection of the stress and frustration detailed by students who needed to prove their innocence. In fact, 27.8\% of the posts in this category explicitly mention Turnitin, 10.2\% mention GPTZero, and 4.6\% mention ZeroGPT.\footnote{These results were calculated by determining the frequency of the tokens “Turnitin,” “GPTZero,” and “ZeroGPT.”} These observations display how AI detector shortcomings can negatively impact innocent students. Despite several studies having already demonstrated the unreliability of AI detectors \cite{elkhatat_evaluating_2023, weber-wulff_testing_2023}, 91.7\% of the posts in this category are recent, from the year 2024. This accentuates an ongoing reliance on fundamentally flawed AI detectors to identify academic integrity violations.


Categories 2 and 3 consist of students who either openly admitted to using AI technology or students who saw other students blatantly abusing AI technology for work in the same class. These two categories show that there is a non-trivial amount of academic misuse of GAI tools from students. Category 4 reflects educators’ reliance on AI detection tools to identify potential academic integrity violations involving GAI. Posts in this category often have negative sentiment, with teachers portraying frustration over students’ misuse of GAI tools. Some educators explicitly voice skepticism about the accuracy of detection tools and acknowledge their limitations. Separately, several teachers mention inadequate institutional support and dissatisfaction with existing school policies on GAI.

\section{Conclusion}

This study contributes to both the educational sector and the advancement of frameworks for analyzing online social discourse. It demonstrates that a prompt-based LLM framework can accurately perform topic modeling, sentiment analysis, and author classification on online discourse (RQ1). The integration of these three tasks also enabled the identification of patterns across author groups that would not emerge from a single-task analysis. While this framework was utilized in the education domain, it is generalizable to other online social systems. 

By modeling the structure of teacher, student, and other stakeholder discourse within online educational communities, this study contributes to the understanding of how important stakeholder groups in education perceive and engage with GAI technologies. For RQ2, there was some alignment across author roles. Both teachers and students are optimistic about GAI’s potential for personalized learning and productivity, and they share uneasiness about GAI misuse. RQ3 was addressed through the topic-specific sentiment analysis, revealing that perceptions varied widely by themes. For instance, teachers emerged as the most vocal group regarding concerns about educator job security. Answering RQ4 revealed underexplored issues through detailed, first-person Reddit accounts, including the emotional toll AI detectors impose on innocent students and educators’ simultaneous reliance on and resistance to these tools.

This study offers valuable perspectives for educators, students, parents, developers, and policymakers. These findings support informed decision-making and the development of balanced policies that respond to the expanding role of GAI in education. For future work, we plan to examine how these topics and sentiments change over time for teachers and students, which will address whether AI is gaining more acceptance in the current LLM era. We also intend to develop a generalized, dataset-agnostic framework for applying and evaluating LLMs for topic modeling and sentiment analysis tasks, as this study demonstrates the effectiveness of a prompt-based LLM for such tasks.


\section*{Data Availability Statement}

\noindent Reddit data utilized in this study were collected via the public Reddit API with PRAW (Python Reddit API Wrapper). The dataset comprises Reddit posts and depth-0 comments. Some posts or comments may have been deleted or altered on the platform after data collection. All analyses were conducted on the dataset as retrieved during the collection window. In accordance with Reddit’s content policy and platform terms, we do not publicly release the full post or comment text. However, we provide post IDs and anonymized metadata upon reasonable request for reproducibility purposes.

\section*{Conflict of Interest}

\noindent There are no conflicts of interest to report.

\section*{Acknowledgment}

\noindent This material is based upon work supported by the National Science Foundation under Grant No. 2332306. 


\begin{thebibliography}{10}
\providecommand{\url}[1]{#1}
\csname url@samestyle\endcsname
\providecommand{\newblock}{\relax}
\providecommand{\bibinfo}[2]{#2}
\providecommand{\BIBentrySTDinterwordspacing}{\spaceskip=0pt\relax}
\providecommand{\BIBentryALTinterwordstretchfactor}{4}
\providecommand{\BIBentryALTinterwordspacing}{\spaceskip=\fontdimen2\font plus
\BIBentryALTinterwordstretchfactor\fontdimen3\font minus \fontdimen4\font\relax}
\providecommand{\BIBforeignlanguage}[2]{{%
\expandafter\ifx\csname l@#1\endcsname\relax
\typeout{** WARNING: IEEEtran.bst: No hyphenation pattern has been}%
\typeout{** loaded for the language `#1'. Using the pattern for}%
\typeout{** the default language instead.}%
\else
\language=\csname l@#1\endcsname
\fi
#2}}
\providecommand{\BIBdecl}{\relax}
\BIBdecl

\bibitem{rosenbaum_microsoft_2024}
\BIBentryALTinterwordspacing
E.~Rosenbaum, ``\BIBforeignlanguage{en}{Microsoft, {Khan} {Academy} provide free {AI} assistant for all educators in {US}},'' May 2024, section: Technology Executive Council. [Online]. Available: \url{https://www.cnbc.com/2024/05/21/microsoft-khan-academy-launch-free-ai-assistant-for-all-us-teachers.html}
\BIBentrySTDinterwordspacing

\bibitem{beatty_khan_2024}
\BIBentryALTinterwordspacing
S.~Beatty, ``\BIBforeignlanguage{en-US}{Khan {Academy} and {Microsoft} partner to expand access to {AI} tools that personalize teaching and help make learning fun},'' May 2024. [Online]. Available: \url{https://news.microsoft.com/source/features/ai/khan-academy-and-microsoft-partner-to-expand-access-to-ai-tools/}
\BIBentrySTDinterwordspacing

\bibitem{noauthor_bring_nodate}
\BIBentryALTinterwordspacing
``\BIBforeignlanguage{en-US}{Bring {AI} to campus at scale}.'' [Online]. Available: \url{https://openai.com/chatgpt/education/}
\BIBentrySTDinterwordspacing

\bibitem{noauthor_arizona_nodate}
\BIBentryALTinterwordspacing
``\BIBforeignlanguage{en-US}{Arizona {State} {University} personalizes learning and advances research with {ChatGPT}}.'' [Online]. Available: \url{https://openai.com/index/asu/}
\BIBentrySTDinterwordspacing

\bibitem{obrien_explainer_2023}
\BIBentryALTinterwordspacing
M.~O'Brien, ``\BIBforeignlanguage{en}{{EXPLAINER}: {What} is {ChatGPT} and why are schools blocking it?}'' Jan. 2023, section: Technology. [Online]. Available: \url{https://apnews.com/article/what-is-chat-gpt-ac4967a4fb41fda31c4d27f015e32660}
\BIBentrySTDinterwordspacing

\bibitem{rosenblatt_new_2023}
\BIBentryALTinterwordspacing
K.~Rosenblatt, ``\BIBforeignlanguage{en}{New {York} {City} public schools remove {ChatGPT} ban},'' May 2023. [Online]. Available: \url{https://www.nbcnews.com/tech/chatgpt-ban-dropped-new-york-city-public-schools-rcna85089}
\BIBentrySTDinterwordspacing

\bibitem{liu_future_2023}
M.~Liu, Y.~Ren, L.~M. Nyagoga, F.~Stonier, Z.~Wu, and L.~Yu, ``\BIBforeignlanguage{en}{Future of education in the era of generative artificial intelligence: {Consensus} among {Chinese} scholars on applications of {ChatGPT} in schools},'' \emph{\BIBforeignlanguage{en}{Future in Educational Research}}, vol.~1, no.~1, pp. 72--101, Sep. 2023.

\bibitem{sinha_how_2023}
\BIBentryALTinterwordspacing
S.~Sinha, L.~Burd, and J.~Du~Perez, ``\BIBforeignlanguage{en}{How {ChatGPT} {Could} {Revolutionize} {Academia} - {IEEE} {Spectrum}},'' Feb. 2023. [Online]. Available: \url{https://spectrum.ieee.org/how-chatgpt-could-revolutionize-academia}
\BIBentrySTDinterwordspacing

\bibitem{chan_ai_2023}
C.~K.~Y. Chan and K.~K.~W. Lee, ``The {AI} generation gap: {Are} {Gen} {Z} students more interested in adopting generative {AI} such as {ChatGPT} in teaching and learning than their {Gen} {X} and millennial generation teachers?'' \emph{Smart Learning Environments}, vol.~10, no.~60, Nov. 2023.

\bibitem{smolansky_educator_2023}
A.~Smolansky, A.~Cram, C.~Raduescu, S.~Zeivots, E.~Huber, and R.~F. Kizilcec, ``Educator and {Student} {Perspectives} on the {Impact} of {Generative} {AI} on {Assessments} in {Higher} {Education},'' in \emph{Proceedings of the {Tenth} {ACM} {Conference} on {Learning} @ {Scale}}, ser. L@{S} '23.\hskip 1em plus 0.5em minus 0.4em\relax New York, NY, USA: Association for Computing Machinery, Jul. 2023, pp. 378--382.

\bibitem{burkhard_student_2022}
\BIBentryALTinterwordspacing
M.~Burkhard, ``\BIBforeignlanguage{en}{Student {Perceptions} of {AI}-{Powered} {Writing} {Tools}: {Towards} {Individualized} {Teaching} {Strategies}},'' International Association for the Development of the Information Society, Tech. Rep., 2022. [Online]. Available: \url{https://eric.ed.gov/?id=ED626893}
\BIBentrySTDinterwordspacing

\bibitem{chan_will_2024}
C.~K.~Y. Chan and L.~H.~Y. Tsi, ``Will generative {AI} replace teachers in higher education? {A} study of teacher and student perceptions,'' \emph{Studies in Educational Evaluation}, vol.~83, p. 101395, Dec. 2024.

\bibitem{vaswani_attention_2017}
A.~Vaswani, N.~Shazeer, N.~Parmar, J.~Uszkoreit, L.~Jones, A.~N. Gomez, L.~Kaiser, and I.~Polosukhin, ``Attention {Is} {All} {You} {Need},'' in \emph{Advances in {Neural} {Information} {Processing} {Systems}}, vol.~30, 2017, pp. 5998--6008.

\bibitem{brown_language_2020}
T.~B. Brown, B.~Mann, N.~Ryder, M.~Subbiah, J.~Kaplan, P.~Dhariwal, A.~Neelakantan, P.~Shyam, G.~Sastry, A.~Askell, S.~Agarwal, A.~Herbert-Voss, G.~Krueger, T.~Henighan, R.~Child, A.~Ramesh, D.~M. Ziegler, J.~Wu, C.~Winter, C.~Hesse, M.~Chen, E.~Sigler, M.~Litwin, S.~Gray, B.~Chess, J.~Clark, C.~Berner, S.~McCandlish, A.~Radford, I.~Sutskever, and D.~Amodei, ``\BIBforeignlanguage{en}{Language {Models} are {Few}-{Shot} {Learners}},'' Jul. 2020, arXiv preprint arXiv:2005.14165.

\bibitem{devlin_bert_2019}
J.~Devlin, M.-W. Chang, K.~Lee, and K.~Toutanova, ``{BERT}: {Pre}-training of {Deep} {Bidirectional} {Transformers} for {Language} {Understanding},'' May 2019, arXiv preprint arXiv:1810.04805.

\bibitem{baidoo-anu_education_2023}
D.~Baidoo-Anu and L.~Owusu~Ansah, ``\BIBforeignlanguage{en}{Education in the {Era} of {Generative} {Artificial} {Intelligence} ({AI}): {Understanding} the {Potential} {Benefits} of {ChatGPT} in {Promoting} {Teaching} and {Learning}},'' \emph{\BIBforeignlanguage{en}{SSRN Electronic Journal}}, 2023.

\bibitem{alasadi_generative_2023}
E.~A. Alasadi and C.~R. Baiz, ``Generative {AI} in {Education} and {Research}: {Opportunities}, {Concerns}, and {Solutions},'' \emph{Journal of Chemical Education}, vol. 100, no.~8, pp. 2965--2971, Aug. 2023, publisher: American Chemical Society.

\bibitem{han_teachers_2024}
A.~Han, X.~Zhou, Z.~Cai, S.~Han, R.~Ko, S.~Corrigan, and K.~A. Peppler, ``Teachers, {Parents}, and {Students}' perspectives on {Integrating} {Generative} {AI} into {Elementary} {Literacy} {Education},'' in \emph{Proceedings of the {CHI} {Conference} on {Human} {Factors} in {Computing} {Systems}}, ser. {CHI} '24.\hskip 1em plus 0.5em minus 0.4em\relax New York, NY, USA: Association for Computing Machinery, May 2024, pp. 1--17.

\bibitem{bull_generative_2023}
C.~Bull and A.~Kharrufa, ``\BIBforeignlanguage{en}{Generative {AI} {Assistants} in {Software} {Development} {Education}},'' Aug. 2023, arXiv preprint arXiv:2303.13936.

\bibitem{rouzegar_generative_2024}
H.~Rouzegar and M.~Makrehchi, ``\BIBforeignlanguage{en}{Generative {AI} for {Enhancing} {Active} {Learning} in {Education}: {A} {Comparative} {Study} of {GPT}-3.5 and {GPT}-4 in {Crafting} {Customized} {Test} {Questions}},'' Jun. 2024, arXiv preprint arXiv:2406.13903.

\bibitem{abbas_is_2024}
M.~Abbas, F.~A. Jam, and T.~I. Khan, ``\BIBforeignlanguage{en}{Is it harmful or helpful? {Examining} the causes and consequences of generative {AI} usage among university students},'' \emph{\BIBforeignlanguage{en}{International Journal of Educational Technology in Higher Education}}, vol.~21, no.~1, p.~10, Feb. 2024.

\bibitem{nazaretsky_teachers_2022}
T.~Nazaretsky, M.~Ariely, M.~Cukurova, and G.~Alexandron, ``\BIBforeignlanguage{en}{Teachers' trust in {AI}-powered educational technology and a professional development program to improve it},'' \emph{\BIBforeignlanguage{en}{British Journal of Educational Technology}}, vol.~53, no.~4, pp. 914--931, Apr. 2022.

\bibitem{chiu_future_2024}
T.~K.~F. Chiu, ``Future research recommendations for transforming higher education with generative {AI},'' \emph{Computers and Education: Artificial Intelligence}, vol.~6, p. 100197, Jun. 2024.

\bibitem{farrokhnia_swot_2024}
M.~Farrokhnia, S.~K. Banihashem, O.~Noroozi, and A.~Wals, ``A {SWOT} analysis of {ChatGPT}: {Implications} for educational practice and research,'' \emph{Innovations in Education and Teaching International}, vol.~61, no.~3, pp. 460--474, May 2024.

\bibitem{cotton_chatting_2024}
D.~R.~E. Cotton, P.~A. Cotton, and J.~R. Shipway, ``Chatting and cheating: {Ensuring} academic integrity in the era of {ChatGPT},'' \emph{Innovations in Education and Teaching International}, vol.~61, no.~2, pp. 228--239, Mar. 2024.

\bibitem{wei_chain--thought_2022}
J.~Wei, X.~Wang, D.~Schuurmans, M.~Bosma, B.~Ichter, F.~Xia, E.~Chi, Q.~V. Le, and D.~Zhou, ``\BIBforeignlanguage{en}{Chain-of-{Thought} {Prompting} {Elicits} {Reasoning} in {Large} {Language} {Models}},'' in \emph{\BIBforeignlanguage{en}{Advances in {Neural} {Information} {Processing} {Systems}}}, vol.~35, Dec. 2022, pp. 24\,824--24\,837.

\bibitem{susnjak_chatgpt_2024}
T.~Susnjak and T.~R. McIntosh, ``\BIBforeignlanguage{en}{{ChatGPT}: {The} {End} of {Online} {Exam} {Integrity}?}'' \emph{\BIBforeignlanguage{en}{Education Sciences}}, vol.~14, no.~6, p. 656, Jun. 2024, number: 6 Publisher: Multidisciplinary Digital Publishing Institute.

\bibitem{fowler_analysis_2023}
\BIBentryALTinterwordspacing
G.~A. Fowler, ``\BIBforeignlanguage{en-US}{Analysis {\textbar} {We} tested a new {ChatGPT}-detector for teachers. {It} flagged an innocent student.}'' \emph{\BIBforeignlanguage{en-US}{Washington Post}}, Apr. 2023. [Online]. Available: \url{https://www.washingtonpost.com/technology/2023/04/01/chatgpt-cheating-detection-turnitin/}
\BIBentrySTDinterwordspacing

\bibitem{ardito_contra_2023}
C.~G. Ardito, ``Contra generative {AI} detection in higher education assessments,'' Dec. 2023, arXiv preprint arXiv:2312.05241.

\bibitem{moorhouse_generative_2023}
B.~L. Moorhouse, M.~A. Yeo, and Y.~Wan, ``Generative {AI} tools and assessment: {Guidelines} of the world's top-ranking universities,'' \emph{Computers and Education}, vol.~5, p. 100151, Dec. 2023.

\bibitem{gocen_artificial_2021}
A.~Gocen and F.~Aydemir, ``\BIBforeignlanguage{en}{Artificial {Intelligence} in {Education} and {Schools}},'' \emph{\BIBforeignlanguage{en}{Research on Education and Media}}, vol.~12, no.~1, pp. 13--21, Jun. 2021.

\bibitem{berendt_ai_2020}
B.~Berendt, A.~Littlejohn, and M.~Blakemore, ``\BIBforeignlanguage{en}{{AI} in {Education}: {Learner} {Choice} and {Fundamental} {Rights}},'' \emph{\BIBforeignlanguage{en}{Learning, Media and Technology}}, vol.~45, no.~3, pp. 312--324, Jun. 2020.

\bibitem{hutto_vader_2014}
C.~Hutto and E.~Gilbert, ``\BIBforeignlanguage{en}{{VADER}: {A} {Parsimonious} {Rule}-{Based} {Model} for {Sentiment} {Analysis} of {Social} {Media} {Text}},'' \emph{\BIBforeignlanguage{en}{Proceedings of the International AAAI Conference on Web and Social Media}}, vol.~8, no.~1, pp. 216--225, May 2014.

\bibitem{noauthor_textblob_nodate}
\BIBentryALTinterwordspacing
``{TextBlob}.'' [Online]. Available: \url{https://textblob.readthedocs.io/en/dev/quickstart.html}
\BIBentrySTDinterwordspacing

\bibitem{wang_is_2024}
Z.~Wang, Q.~Xie, Y.~Feng, Z.~Ding, Z.~Yang, and R.~Xia, ``Is {ChatGPT} a {Good} {Sentiment} {Analyzer}? {A} {Preliminary} {Study},'' Feb. 2024, arXiv preprint arXiv:2304.04339.

\bibitem{zhang_sentiment_2023}
W.~Zhang, Y.~Deng, B.~Liu, S.~J. Pan, and L.~Bing, ``Sentiment {Analysis} in the {Era} of {Large} {Language} {Models}: {A} {Reality} {Check},'' May 2023, arXiv preprint arXiv:2305.15005.

\bibitem{kheiri_sentimentgpt_2023}
K.~Kheiri and H.~Karimi, ``{SentimentGPT}: {Exploiting} {GPT} for {Advanced} {Sentiment} {Analysis} and its {Departure} from {Current} {Machine} {Learning},'' Jul. 2023, arXiv preprint arXiv:2307.10234.

\bibitem{rosenthal_semeval-2017_2019}
S.~Rosenthal, N.~Farra, and P.~Nakov, ``{SemEval}-2017 {Task} 4: {Sentiment} {Analysis} in {Twitter},'' Dec. 2019, arXiv preprint arXiv:1912.00741.

\bibitem{blei_latent_2003}
D.~M. Blei, A.~Y. Ng, and M.~I. Jordan, ``\BIBforeignlanguage{en}{Latent {Dirichlet} {Allocation}},'' \emph{\BIBforeignlanguage{en}{Journal of Machine Learning Research}}, vol.~3, pp. 993--1022, Jan. 2003.

\bibitem{tong_text_2016}
Z.~Tong and H.~Zhang, ``\BIBforeignlanguage{en}{A {Text} {Mining} {Research} {Based} on {LDA} {Topic} {Modelling}},'' in \emph{\BIBforeignlanguage{en}{Computer {Science} \& {Information} {Technology} ( {CS} \& {IT} )}}.\hskip 1em plus 0.5em minus 0.4em\relax Academy \& Industry Research Collaboration Center (AIRCC), May 2016, pp. 201--210.

\bibitem{grootendorst_bertopic_2022}
M.~Grootendorst, ``{BERTopic}: {Neural} topic modeling with a class-based {TF}-{IDF} procedure,'' Mar. 2022, arXiv preprint arXiv:2203.05794.

\bibitem{angelov_top2vec_2020}
D.~Angelov, ``{Top2Vec}: {Distributed} {Representations} of {Topics},'' Aug. 2020, arXiv preprint arXiv:2008.09470.

\bibitem{phamTopicGPTPromptbasedTopic2024}
C.~M. Pham, A.~Hoyle, S.~Sun, P.~Resnik, and M.~Iyyer, ``{TopicGPT}: {A} {Prompt}-based {Topic} {Modeling} {Framework},'' Apr. 2024, arXiv preprint arXiv:2311.01449.

\bibitem{wang_prompting_2023}
H.~Wang, N.~Prakash, N.~K. Hoang, M.~S. Hee, U.~Naseem, and R.~K.-W. Lee, ``Prompting {Large} {Language} {Models} for {Topic} {Modeling},'' in \emph{2023 {IEEE} {International} {Conference} on {Big} {Data} ({BigData})}, Dec. 2023, pp. 1236--1241.

\bibitem{mu_large_2024}
Y.~Mu, C.~Dong, K.~Bontcheva, and X.~Song, ``Large {Language} {Models} {Offer} an {Alternative} to the {Traditional} {Approach} of {Topic} {Modelling},'' Mar. 2024, arXiv preprint arXiv:2403.16248.

\bibitem{li_sentiment_2023}
S.~Li, Z.~Xie, D.~K.~W. Chiu, and K.~K.~W. Ho, ``\BIBforeignlanguage{en}{Sentiment {Analysis} and {Topic} {Modeling} {Regarding} {Online} {Classes} on the {Reddit} {Platform}: {Educators} versus {Learners}},'' \emph{\BIBforeignlanguage{en}{Applied Sciences}}, vol.~13, no.~4, p. 2250, Jan. 2023.

\bibitem{mujahid_sentiment_2021}
M.~Mujahid, E.~Lee, F.~Rustam, P.~B. Washington, S.~Ullah, A.~A. Reshi, and I.~Ashraf, ``\BIBforeignlanguage{en}{Sentiment {Analysis} and {Topic} {Modeling} on {Tweets} about {Online} {Education} during {COVID}-19},'' \emph{\BIBforeignlanguage{en}{Applied Sciences}}, vol.~11, no.~18, p. 8438, Jan. 2021.

\bibitem{choi_exploring_2023}
W.~Choi, Y.~Zhang, and B.~Stvilia, ``\BIBforeignlanguage{en}{Exploring {Applications} and {User} {Experience} with {Generative} {AI} {Tools}: {A} {Content} {Analysis} of {Reddit} {Posts} on {ChatGPT}},'' \emph{\BIBforeignlanguage{en}{Proceedings of the Association for Information Science and Technology}}, vol.~60, no.~1, pp. 543--546, 2023.

\bibitem{naing_public_2024}
S.~Z.~S. Naing and P.~Udomwong, ``Public {Opinions} on {ChatGPT} : {An} {Analysis} of {Reddit} {Discussions} by {Using} {Sentiment} {Analysis}, {Topic} {Modeling}, and {SWOT} {Analysis},'' \emph{Data Intelligence}, vol.~6, no.~2, pp. 344--374, May 2024.

\bibitem{futterer_chatgpt_2023}
T.~Fütterer, C.~Fischer, A.~Alekseeva, X.~Chen, T.~Tate, M.~Warschauer, and P.~Gerjets, ``\BIBforeignlanguage{en}{{ChatGPT} in education: global reactions to {AI} innovations},'' \emph{\BIBforeignlanguage{en}{Scientific Reports}}, vol.~13, no.~1, p. 15310, Sep. 2023.

\bibitem{kuhaneswaran_exploring_2024}
B.~Kuhaneswaran, A.~Vadivel, A.~Wijeratne, N.~Ravikumar, S.~Kumara, A.~Sivapalan, and T.~Vijayanandan, ``\BIBforeignlanguage{en}{Exploring the {Educational} {Landscape} of {ChatGPT}: {A} {Topic} {Modeling} {Approach} on {Twitter} {Data}},'' \emph{\BIBforeignlanguage{en}{Sri Lanka Journal of Social Sciences and Humanities}}, vol.~4, no.~1, pp. 1--12, Oct. 2024.

\bibitem{smith-mutegi_perceptions_2025}
D.~Smith-Mutegi, Y.~Mamo, J.~Kim, H.~Crompton, and M.~McConnell, ``Perceptions of {STEM} education and artificial intelligence: a {Twitter} ({X}) sentiment analysis,'' \emph{International Journal of STEM Education}, vol.~12, no.~1, p.~9, Feb. 2025.

\bibitem{mamo_higher_2024}
Y.~Mamo, H.~Crompton, D.~Burke, and C.~Nickel, ``\BIBforeignlanguage{en}{Higher {Education} {Faculty} {Perceptions} of {ChatGPT} and the {Influencing} {Factors}: {A} {Sentiment} {Analysis} of {X}},'' \emph{\BIBforeignlanguage{en}{TechTrends}}, vol.~68, no.~3, pp. 520--534, May 2024.

\bibitem{lee_community_2024}
Y.~Lee, ``\BIBforeignlanguage{en}{Community {Perspectives} on {ChatGPT}: {Sentiment} {Analysis} in {Educational} {Forum}},'' \emph{\BIBforeignlanguage{en}{TechTrends}}, vol.~68, no.~6, pp. 1195--1207, Nov. 2024.

\bibitem{oh_explore_2024}
S.~Oh, Y.~Cao, A.~Katz, and J.~Zhao, ``Explore {Public}'s {Perspectives} on {Generative} {AI} in {Computer} {Science} ({CS}) {Education}: {A} {Social} {Media} {Data} {Analysis},'' in \emph{2024 {IEEE} {Frontiers} in {Education} {Conference} ({FIE})}, Oct. 2024, pp. 1--9.

\bibitem{noauthor_praw_nodate}
\BIBentryALTinterwordspacing
``{PRAW} 7.7.1 documentation.'' [Online]. Available: \url{https://praw.readthedocs.io/en/stable/}
\BIBentrySTDinterwordspacing

\bibitem{rehurek_software_2010}
R.~Řehůřek and P.~Sojka, ``Software {Framework} for {Topic} {Modelling} with {Large} {Corpora},'' in \emph{Proceedings of the {LREC} 2010 {Workshop} on {New} {Challenges} for {NLP} {Frameworks}}.\hskip 1em plus 0.5em minus 0.4em\relax ELRA, May 2010, pp. 45--50.

\bibitem{khattab_dspy_2023}
O.~Khattab, A.~Singhvi, P.~Maheshwari, Z.~Zhang, K.~Santhanam, S.~Vardhamanan, S.~Haq, A.~Sharma, T.~T. Joshi, H.~Moazam, H.~Miller, M.~Zaharia, and C.~Potts, ``{DSPy}: {Compiling} {Declarative} {Language} {Model} {Calls} into {Self}-{Improving} {Pipelines},'' Oct. 2023, arXiv preprint arXiv:2310.03714.

\bibitem{noauthor_all-minilm-l6-v2_2024}
\BIBentryALTinterwordspacing
``all-{MiniLM}-{L6}-v2,'' Jan. 2024. [Online]. Available: \url{https://huggingface.co/sentence-transformers/all-MiniLM-L6-v2}
\BIBentrySTDinterwordspacing

\bibitem{noauthor_all-mpnet-base-v2_2024}
\BIBentryALTinterwordspacing
``all-mpnet-base-v2,'' Jan. 2024. [Online]. Available: \url{https://huggingface.co/sentence-transformers/all-mpnet-base-v2}
\BIBentrySTDinterwordspacing

\bibitem{elkhatat_evaluating_2023}
A.~M. Elkhatat, K.~Elsaid, and S.~Almeer, ``\BIBforeignlanguage{en}{Evaluating the efficacy of {AI} content detection tools in differentiating between human and {AI}-generated text},'' \emph{\BIBforeignlanguage{en}{International Journal for Educational Integrity}}, vol.~19, no.~1, p.~17, Sep. 2023.

\bibitem{weber-wulff_testing_2023}
D.~Weber-Wulff, A.~Anohina-Naumeca, S.~Bjelobaba, T.~Foltýnek, J.~Guerrero-Dib, O.~Popoola, P.~Šigut, and L.~Waddington, ``\BIBforeignlanguage{en}{Testing of detection tools for {AI}-generated text},'' \emph{\BIBforeignlanguage{en}{International Journal for Educational Integrity}}, vol.~19, no.~1, p.~26, Dec. 2023.

\end{thebibliography}

\end{document}